\documentclass[10pt,letterpaper]{article}
\usepackage[top=0.85in,left=2.75in,footskip=0.75in]{geometry}

\usepackage{amsmath,amssymb}
\usepackage{multirow}

\usepackage{changepage}

\usepackage{textcomp,marvosym}

\usepackage{cite}

\usepackage{nameref,hyperref}

\usepackage{chngcntr}

\usepackage[nopatch=eqnum]{microtype}
\DisableLigatures[f]{encoding = *, family = * }

\usepackage[table]{xcolor}

\usepackage{array}

\newcolumntype{+}{!{\vrule width 2pt}}

\newlength\savedwidth

\raggedright
\usepackage[aboveskip=1pt,labelfont=bf,labelsep=period,justification=raggedright,singlelinecheck=off]{caption}

\makeatletter
\renewcommand{\@biblabel}[1]{\quad#1.}
\makeatother

\usepackage{lastpage,fancyhdr,graphicx}
\usepackage{epstopdf}
\fancyheadoffset[L]{2.25in}
\fancyfootoffset[L]{2.25in}
\begin{document}
\vspace*{0.2in}

\begin{flushleft}
{\Large
\textbf{Coevolutionary dynamics of cooperation, risk, and cost in collective risk games} 
}
\newline
\\
Lichen Wang\textsuperscript{1\Yinyang},
Shijia Hua\textsuperscript{1\Yinyang},
Yuyuan Liu\textsuperscript{1},
Liang Zhang\textsuperscript{1},
Linjie Liu\textsuperscript{1,2*},
Attila Szolnoki\textsuperscript{3},
\\
\bigskip
\textbf{1} College of Science, Northwest A \& F University, Yangling, Shaanxi, China
\\
\textbf{2} College of Economics \& Management, Northwest A \& F University, Yangling, Shaanxi, China
\\
\textbf{3} Institute of Technical Physics and Materials Science, Centre for Energy Research, Budapest, Hungary
\\
\bigskip

%
%
\Yinyang These authors contributed equally to this work.





* linjie140@126.com

\end{flushleft}


%
\section*{Abstract}
Addressing both natural and societal challenges requires collective cooperation. Studies on collective-risk social dilemmas have shown that individual decisions are influenced by the perceived risk of collective failure. However, existing feedback-evolving game models often focus on a single feedback mechanism, such as the coupling between cooperation and risk or between cooperation and cost. In many real-world scenarios, however, the level of cooperation, the cost of cooperating, and the collective risk are dynamically interlinked. Here, we present an evolutionary game model that considers the interplay of these three variables. Our analysis shows that the worst-case scenario, characterized by full defection, maximum risk, and the highest cost of cooperation\textcolor{red}{,} remains a stable evolutionary attractor. Nevertheless, cooperation can emerge and persist because the system also supports stable equilibria with non-zero cooperation. The system exhibits multistability, meaning that different initial conditions lead to either sustained cooperation or a tragedy of the commons. These findings highlight that initial levels of cooperation, cost, and risk collectively determine whether a population can avert a tragic outcome.

\section*{Author summary}
Collective cooperation is indispensable for addressing natural and social challenges. Prior research on collective-risk social dilemmas has shown that individual contributions are shaped by the risk of collective failure, yet it insufficiently captures a key complexity: the intricate interdependencies among cooperation cost, collective failure risk, and cooperation level. Here, we develop an evolutionary game model with a multiple feedback mechanism to clarify cooperative behavior emergence in complex gaming contexts. We find that this tightly coupled system yields diverse dynamic outcomes. Notably, individuals can sustain stable cooperation in low-risk environments with only minimal cooperation costs. However, the ``tragedy of the commons", a state of universal free-riding with maximum risk and cooperation costs is always attainable in our model. Critical factors for escaping this dilemma include the initial cooperation rate, cooperation cost, and risk value.


\section*{Introduction}
The interaction between humans and their surrounding environment constitutes a complex system, in which changes in one profoundly affect the other~\cite{mysterud2006concept,liu2007complexity,kalantari2019meeting,farahbakhsh2022modelling,gret2019actors}. Current human activities such as urbanization, agricultural expansion, and industrial production not only change the structure and function of natural ecosystems but also have significant impacts on biodiversity, climate change, and land cover~\cite{granot1998dark,tilman1999global,patz2005impact,moore2009climate,laurance2014agricultural,obradovich2019risk,goossens2021air}. In turn, these environmental changes also shape human lifestyles, health conditions, and socio-economic development. For example, extreme weather events caused by climate change and the reduction of natural resources have begun to affect global food security and water security, further exacerbating social inequality and migration issues~\cite{frederick1997climate,gregory2005climate,schuur2015climate,el2020climate}. Therefore, understanding the interaction mechanism between humans and the environment is crucial for formulating sustainable development strategies and addressing the challenges posed by environmental change~\cite{perc2010coevolutionary,wang2020steering,kleshnina2023effect,su2019evolutionary,wang2024evolution}.

The dynamic interplay between human behavior and the environment can be conceptualized as a complex coupled game model, in which the feedback-evolving game has recently attracted significant research attention. This pioneering framework of coevolutionary game theory was proposed by Weitz {\it et al.}~\cite{weitz2016oscillating} who considered that individuals' strategy choices and behaviors are not only influenced by their own and opponent's strategies but also by direct or indirect feedback from their changing environment. Along this line, feedback-evolving game models have led to a rich body of literature, including classic public goods games, collective-risk social dilemma games, as well as studies on common-pool resources and adaptive incentives~\cite{kurokawa2009emergence,kraak2011exploring,wang2024paradigm,hua2024coevolutionary1,wang2023emergence,wang2020eco}. Among these, the collective-risk social dilemma with feedback mechanisms is particularly relevant, as it captures the critical feedback between behavioral decisions and environmental fragility in addressing global challenges like climate change and epidemic control~\cite{hua2024coevolutionary2,liu2023coevolutionary}.

To understand outcomes in such dilemmas, it is essential to examine the roles of risk perception and cooperation in individual and collective decision-making. Evolutionary game theory serves as a robust analytical framework for examining these interconnected systems~\cite{Smith_1982,schuster1983replicator,nowak2004evolutionary,niehus2021evolution,sigmund2021toward}. Previous research has acknowledged that the level of collective risk significantly shapes individual choices, with higher risk generally promoting cooperation to avert collective failure~\cite{milinski2008collective,santos2011risk}.

However, the majority of earlier research treats both risk and the cost of cooperation as static parameters. This approach overlooks a key reality: risk is not static but is subject to dynamic shaping by human behavior~\cite{liu2023coevolutionary}. It has been demonstrated that highly cooperative societies often mitigate collective risks more effectively. This phenomenon can be observed in various contexts, including the reduction of carbon emissions to mitigate global climate change and the enhancement of vaccination rates to curb the transmission of diseases~\cite{chen2022highly,pacheco2024co}. This bidirectional feedback between individual decisions and collective risk has attracted attention from both empirical and theoretical sides. Furthermore, the cost of cooperation is rarely fixed. It is often context-dependent, influenced by both the prevailing risk and the level of cooperation within a group. For example, in a highly cooperative group, the per-capita burden of contribution can be lower. Conversely, a minority of cooperators in a defecting group may face disproportionately high costs, further deterring cooperation~\cite{hua2024coevolutionary2}. To give an illustration from climate change, substantial global collaboration enables the development and implementation of pertinent technologies, thereby reducing the individual nations' emission reduction costs. Conversely, insufficient cooperation among nations results in the imposition of greater economic and environmental burdens on individual countries~\cite{vasconcelos2014climate,hagel2016risk}. Behavioral experiments also indicate that individuals are willing to bear higher costs when facing greater collective risks~\cite{wang2020communicating}.

Despite the growing emphasis on such feedback mechanisms, existing research has yet to fully elucidate the complex co-evolutionary relationship between human decision-making, the costs of cooperation, and collective risks. Here, we develop a model that explicitly couples the dynamics of these three variables, as schematically summarized in Fig~\ref{fig: model_1_figure12}. Our analysis reveals that the tragedy of the commons state, characterized by full defection, maximal risk, and maximal cooperation cost, persists as a stable evolutionary outcome. However, the system also exhibits bistable and tristable evolutionary outcomes, which provide conditions for escaping the aforementioned undesired destination. Our theoretical predictions are further supported by numerical simulations.

\section*{Materials and methods}
We randomly select $N$ individuals from an infinite population to participate in a collective-risk social dilemma game, where each individual's initial endowment is $b$~\cite{milinski2006stabilizing,milinski2008collective,hilbe2013evolution}. Participants have two options: to cooperate by contributing $c$ to a common pool, or to defect without contributing. Each participant retains their residual endowment if the number of cooperators in the group reaches the threshold $M$. Otherwise, participants lose their entire endowment with probability $r$ that characterizes the risk level of collective failure~\cite{milinski2006stabilizing,milinski2008collective,santos2011risk}. Let $N_C$ denote the number of cooperators and $N-N_C$ denote the number of defectors. The payoffs for cooperators and defectors are as follows:
\begin{align}
		P_{C}(N_C)&=b \chi(N_{C}-M)+(1-r) b\left[1-\chi(N_{C}-M)\right]-c\,, \label{eq1}\\
		P_{D}(N_C)&=b \chi(N_{C}-M)+(1-r) b\left[1-\chi(N_{C}-M)\right]\,,\label{eq11}
\end{align}
where $\chi(\xi)$ is the step function defined as $\chi(\xi)=0$ when $\xi<0$ and $\chi(\xi)=1$ when $\xi\geq 0$.

Let $x$ denote the fraction of cooperators within the population. The time evolution of cooperation is described by the replicator equation~\cite{schuster1983replicator}:
\begin{eqnarray}\label{eq2}
\dot{x} = x(1-x)\left(f_{C} - f_{D}\right),
\end{eqnarray}
where $f_C$ and $f_D$ denote the average fitness of cooperators and defectors, respectively. Here, $N_C$ denotes the number of cooperators among the other $N-1$ group members, so that $N_C+1$ accounts for the focal cooperator itself. 
These fitness values are calculated as follows:

\begin{align}
	f_C &= \sum_{N_C=0}^{N-1} \binom{N-1}{N_C} x^{N_C} (1-x)^{N-1-N_C} P_C(N_C + 1),\label{eq3} \\ 
	f_D &= \sum_{N_C=0}^{N-1} \binom{N-1}{N_C} x^{N_C} (1-x)^{N-1-N_C} P_D(N_C), \label{eq4}
\end{align}
where $\binom{N-1}{N_C}$ denotes the binomial coefficient representing the number of ways to choose $N_C$ cooperators from the remaining $N-1$ participants. 

By combining Eqs.~(\ref{eq1})-(\ref{eq4}), we derive an equation that describes how the fraction of cooperators evolves over time: 
\begin{eqnarray} 
\dot{x}=x(1-x)\left[\binom{N-1}{M-1} x^{M-1}(1-x)^{N-M} r b-c\right].
\end{eqnarray}

\begin{figure*}[t!]
\centering
\includegraphics[width=\linewidth]{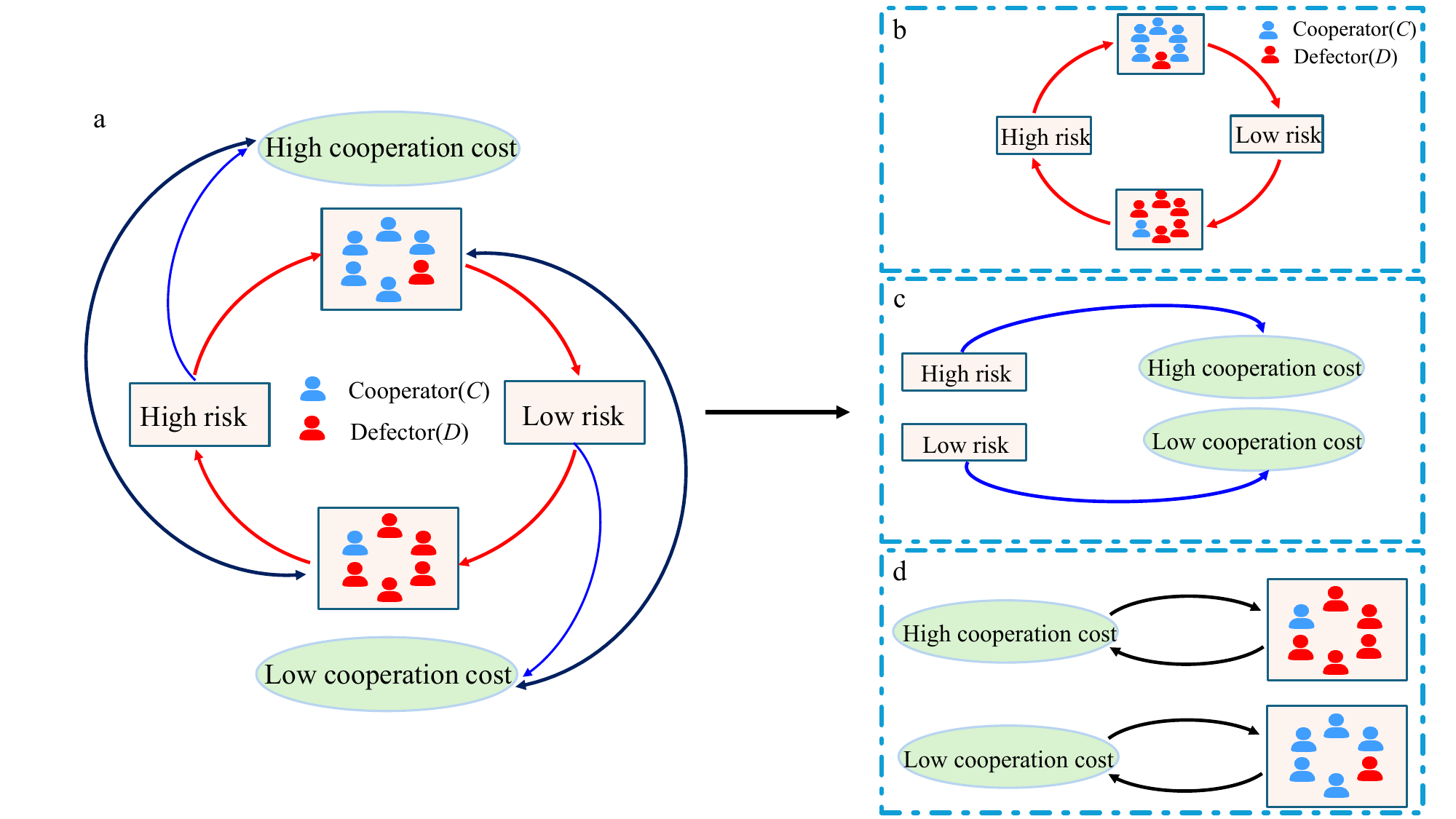}
\caption{{\bf Coevolution of cooperation level, cooperation cost, and risk.} An overview of their subtle interactions is shown in panel~a. The blue line illustrates the dynamic interplay between cooperation cost and collective risk, highlighting their bidirectional influence within the system. The red curve with arrows illustrates the bidirectional coupling between group states and collective risk. The black bidirectional arrows indicate the reciprocal interaction between group states and cooperation cost. The feedback relationship between group states and collective risk is shown in panel~b: an increase in risk promotes cooperative behavior, and the resulting higher level of cooperation reduces risk. However, this reduction may subsequently encourage defection, leading to an increase in risk again. Panel~c shows that high risk increases the cooperation cost, whereas lower risk allows the cooperation cost to decrease. Panel~d illustrates that high cooperation cost makes defection more attractive, which may result in an even higher cooperation cost. Conversely, low cooperation cost makes cooperation more accessible, and the resulting increase in cooperation level further reduces the cooperation cost.}
\label{fig: model_1_figure12}
\end{figure*}

To go beyond and generalize previous studies \cite{chen2012risk,milinski2016humans,domingos2020timing,vasconcelos2013bottom,vasconcelos2014climate}, we consider a bidirectional feedback relationship among the individual cooperation cost $c$, the risk level $r$ and the cooperation level $x$. Motivated by this interplay, we devise a multi-feedback game model as follows:
\begin{equation}\label{eq5}
\begin{cases}
\varepsilon_1 \dot{x}=x(1-x)\left[\binom{N-1}{M-1} x^{M-1}(1-x)^{N-M} r b-c\right]\,, \\
\varepsilon_2 \dot{r}=U_1(x, r)\,,\\
\dot{c}=U_2(x,r,c),
\end{cases}
\end{equation}
where $\varepsilon_1$ and $\varepsilon_2$ represent the relative speeds of strategy and risk updates, respectively~\cite{weitz2016oscillating}. For simplicity, we set $\varepsilon_1=\varepsilon_2=1$. In the S1 File, however, we systematically explore the effects of their variations on the evolutionary outcomes ( Figs A and G in S1 File). The functions $U_1(x,r)$ and $U_2(x,r,c)$ represent the interactions between strategy, risk, and individual contributions. We assume that an increase in the fraction of defectors elevates the collective risk level and the expected cooperation cost, whereas a higher fraction of cooperators reduces both risk and cooperation cost. In addition, higher collective risk is assumed to increase the cooperation cost, while higher cooperation cost discourages cooperation (see Fig~\ref{fig: model_1_figure12}). For simplicity, we assume that these effects are linear, expressed as:
\begin{align}
	U_1(x, r)&=r(1-r)\left[\mu_1(1-x)-\mu_2x\right],\label{eq6}\\
	U_2(x,r,c)&=(\alpha-c)(c-\beta)\left[\theta_1(1-x)-\theta_2x+\theta_3r-\theta_4(1-r)\right]\label{eq66},
\end{align}
where $\mu_1$, $\mu_2$, $\theta_1$, $\theta_2$, $\theta_3$, and $\theta_4$ act as linear coefficients, all of which are positive numbers. Additionally, $\alpha$ and $\beta$ define the upper and lower bounds for the cost $c$, respectively, with $\alpha > \beta > 0$.

By combining Eqs.~(\ref{eq5}), (\ref{eq6}), and (\ref{eq66}), we obtain the following multi-feedback dynamical system:
\begin{equation}\label{eq7}
	\begin{cases}
		 \dot{x} = x(1-x)\left[\binom{N-1}{M-1} x^{M-1}(1-x)^{N-M} r b - c\right], \\
		 \dot{r} = r(1-r)\left[\mu_1(1-x) - \mu_2 x\right], \\
		\dot{c} = (\alpha - c)(c - \beta)\left[\theta_1(1-x) - \theta_2 x + \theta_3 r - \theta_4(1-r)\right].
	\end{cases}
\end{equation}

In order to assess the effectiveness of cooperation in collective social-risk dilemmas, we further define the success rate of cooperation in the collective-risk social dilemma. For a fixed threshold $M$, if the fraction of cooperators in the population stabilizes at $p$, the success rate is given by
\begin{equation*}
    P_{CS} = 1 - \sum_{k=0}^{M-1}\binom{N}{k}p^k(1-p)^{N-k}.
\end{equation*}

To further characterize the system, we theoretically analyze all the equilibrium points of this coupled system and present the stability conditions in the S1 File.
Additionally, we provide a variant of our model in which the cooperation cost also affects the risk level (Fig C in S1 File). 


\section*{Results}

\begin{figure*}[t!]
\centering
\includegraphics[width=\linewidth]{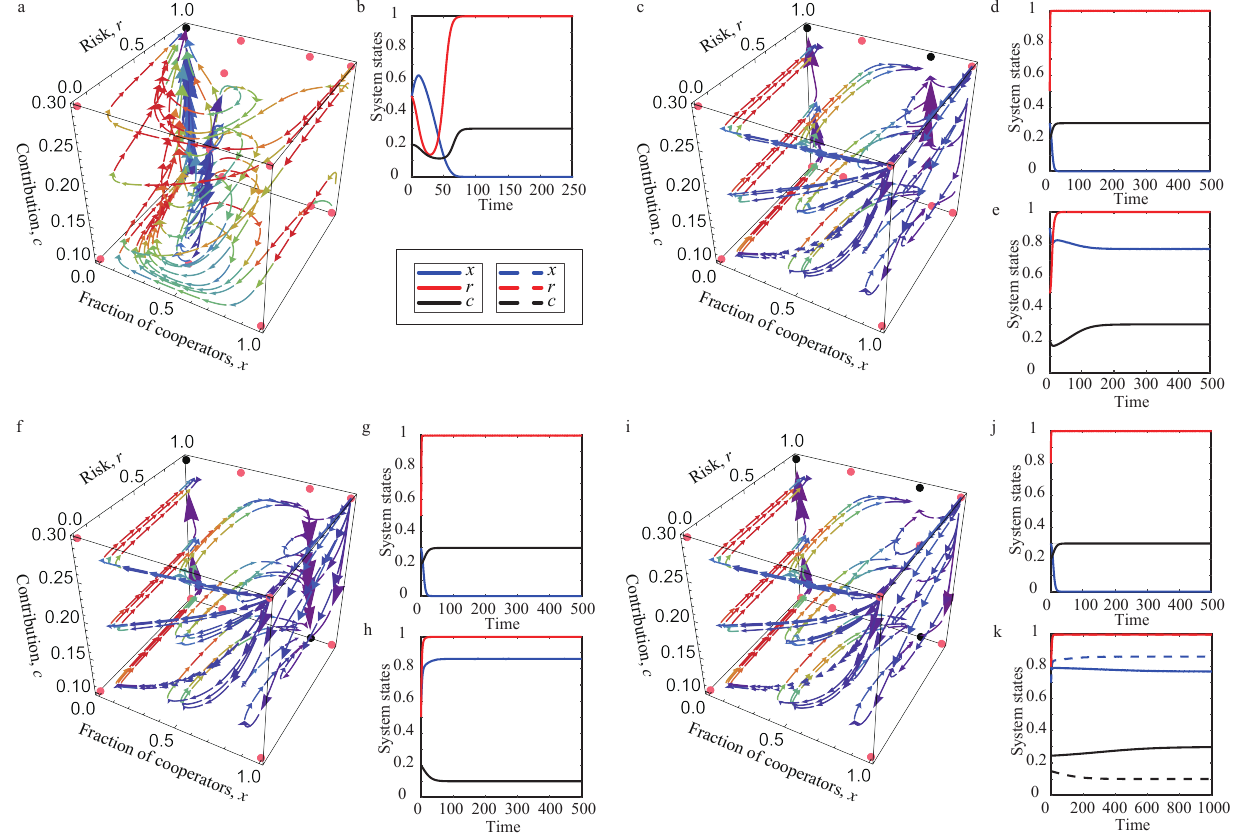}
\caption{{\bf Replicator dynamics with stable corner and edge equilibrium points.} Panels~a, c, f, and i show the phase flow in the $x$–$r$–$c$ phase space, illustrating four distinct evolutionary outcomes. Red points represent unstable equilibrium points, while black points indicate stable equilibrium points. Panels~b, d, e, g, h, j, and k present the time evolution of the coevolving variables. Parameters are $\mu_1=0.25$, $\theta_2=0.5$, $\theta_3=0.6$ in panels~a and b; $\mu_1=2.5$, $\theta_2=0.7$, $\theta_3=0.6$ in panels~c, d, and e; $\mu_1=2.5$, $\theta_2=0.7$, $\theta_3=0.15$ in panels~f, g, and h; $\mu_1=2.5$, $\theta_2=0.5$, $\theta_3=0.3$ in panels~i, j, and k. The remaining parameters $N=8$, $M=5$, $\mu_2=0.3$, $\theta_1=0.5$, $\theta_4=0.6$, $b=2$, $\beta=0.1$, and $\alpha=0.3$ are fixed across all panels.}
\label{fig: model_1_figure1}
\end{figure*}

The feedback-evolving game model presented in Eq.~\eqref{eq7} admits multiple equilibrium solutions (see S1 File for the labeling convention of equilibrium points). In particular, we identify eight corner equilibria, namely $\left(0,0,\alpha\right)$, $\left(0,0,\beta\right)$, $\left(0,1,\alpha\right)$, $\left(0,1,\beta\right)$, $\left(1,0,\alpha\right)$, $\left(1,0,\beta\right)$, $\left(1,1,\alpha\right)$, and $\left(1,1,\beta\right)$. Furthermore, if $b\left(\frac{N-M}{N-1}\right)^{N-M}\left(\frac{M-1}{N-1}\right)^{M-1}\binom{N-1}{M-1}>\alpha$ holds, the system possesses two edge equilibrium points $\left(x_{1t},1,\alpha\right)$ and $\left(x_{2t},1,\alpha\right)$ where $x_{1t}$ and $x_{2t}$ represent the two distinct roots of the equation $b\binom{N-1}{M-1}x^{M-1}(1-x)^{N-M}=\alpha$ and satisfy $x_{1t}<x_{2t}$. Similar conditions determine the existence of the edge equilibrium points $\left(x_{1b},1,\beta\right)$ and $\left(x_{2b},1,\beta\right)$ if $b\left(\frac{N-M}{N-1}\right)^{N-M}\left(\frac{M-1}{N-1}\right)^{M-1}\binom{N-1}{M-1}>\beta$ is satisfied, with $x_{1b}$ and $x_{2b}$ being the two distinct roots of $b\binom{N-1}{M-1}x^{M-1}(1-x)^{N-M}=\beta$ and $x_{1b}<x_{2b}$. Moreover, when $0<B_r<1$, the system has a surface equilibrium point $\left(x^\ast,B_r,\beta\right)$, where $x^\ast = \frac{\mu_1}{\mu_1+\mu_2}$ and $B_r = \frac{\beta}{b\binom{N-1}{M-1} {x^\ast}^{M-1}\left(1-x^\ast\right)^{N-M}}$. When $\theta_3<\theta_2$ and $\beta<R_c<\alpha$, the system has a surface equilibrium point $\left(R_x,1,R_c\right)$, where $R_x = \frac{\theta_1+\theta_3}{\theta_1+\theta_2}$ and $R_c=b\binom{N-1}{M-1}{R_x}^{M-1}\left(1-R_x\right)^{N-M}$. Moreover, when $0<T_r<1$, the system has a surface equilibrium point $\left(x^\ast,T_r,\alpha\right)$, where $T_r = \frac{\alpha}{b\binom{N-1}{M-1} {x^\ast}^{M-1}\left(1-x^\ast\right)^{N-M}}$. When $0<r^\ast<1$ and $\beta<c^\ast<\alpha$, the system admits an interior equilibrium point $\left(x^\ast,r^\ast,c^\ast\right)$, where $r^\ast=\frac{\theta_2\mu_1-\theta_1\mu_2+\theta_4(\mu_1+\mu_2)}{(\theta_3+\theta_4)(\mu_1+\mu_2)}$ and $c^\ast=\binom{N-1}{M-1}{x^\ast}^{M-1}(1-x^\ast)^{N-M}r^\ast b$.

\begin{table}[t!]
	\centering
		\caption{Stability conditions for all equilibrium points of system~\eqref{eq7}.}
	
	\begin{tabular}{ll}
		\hline
		\textbf{Equilibrium $(x, r, c)$} & \textbf{Stability condition} \\
		\hline
		$(0, 1, \alpha)$ & Stable \\
		
		$(0,0,\alpha)$, $(0,0,\beta)$, $(0,1,\beta)$, & \multirow{2}{*}{Unstable} \\
		$(1,0,\alpha)$, $(1,0,\beta)$, $(1,1,\alpha)$, $(1,1,\beta)$ & \\
		
		$(x_{1t}, 1, \alpha)$ 
		& Unstable  \\
		
		$(x_{2t}, 1, \alpha)$ 
		& Stable if $  x_{2t} < \min\{\frac{\mu_1}{\mu_1+\mu_2},\,\frac{\theta_1+\theta_3}{\theta_1+\theta_2}\}$  \\
		
		$(x_{1b}, 1, \beta)$ 
		& Unstable  \\
		
		$(x_{2b}, 1, \beta)$ 
		& Stable if $ \frac{\theta_1+\theta_3}{\theta_1+\theta_2} < x_{2b} < \frac{\mu_1}{\mu_1+\mu_2}$  \\
		
		$(R_x, 1, R_c)$ & Unstable \\
		
		$(x^\ast, T_r, \alpha)$ & Stable if $T_1<0$ and $T_5<0$ \\
		
		$(x^\ast, B_r, \beta)$ & Stable if $B_1<0$ and $B_5<0$ \\
		
		$(x^\ast, r^\ast, c^\ast)$ & Unstable \\
		\hline
	\end{tabular}
	\label{tab:equilibria_full}
\end{table}
All equilibrium points of system~\eqref{eq7} and their corresponding stability conditions are summarized in Table~\ref{tab:equilibria_full}. The detailed derivation of these equilibria and a full stability analysis are provided in the S1 File.
The auxiliary variables $T_1$, $T_5$, $B_1$, and $B_5$ are defined as follows:
	\begin{align*}
		T_1 &= \alpha(M - Nx^\ast - x^\ast) + \alpha(2x^\ast - 1), \\
		T_5 &= (\beta - \alpha)\left[\theta_1(1 - x^\ast) - \theta_2 x^\ast + \theta_3 T_r - \theta_4(1 - T_r)\right], \\
		B_1 &= \beta(M - Nx^\ast - x^\ast) + \beta(2x^\ast - 1), \\
		B_5 &= (\alpha - \beta)\left[\theta_1(1 - x^\ast) - \theta_2 x^\ast + \theta_3 B_r - \theta_4(1 - B_r)\right].
\end{align*}

For corner equilibria, only the equilibrium point $(0, 1, \alpha)$ is stable, whereas all other corner equilibria are unstable. Numerical evidence supports these findings. This is illustrated in Fig~\ref{fig: model_1_figure1}a and \ref{fig: model_1_figure1}b, where black dots indicate stable equilibria and red dots denote unstable equilibria. This outcome corresponds to a worst-case scenario in which defectors prevail under maximal risk, leaving all participants without any payoff.

The detrimental situation becomes avoidable once the edge equilibria $(x_{1t}, 1, \alpha)$ and $(x_{2t}, 1, \alpha)$ exist. The edge equilibrium $(x_{2t}, 1, \alpha)$ is stable if $x_{2t} < \min\left\{\frac{\mu_1}{\mu_1 + \mu_2}, \frac{\theta_1 + \theta_3}{\theta_1 + \theta_2}\right\}$. The supporting numerical results are presented in Fig~\ref{fig: model_1_figure1}c, \ref{fig: model_1_figure1}d and \ref{fig: model_1_figure1}e. These plots demonstrate that, due to bistability, the system converges to different equilibrium points depending on the initial conditions. Notably, the equilibrium point $\left(0,1,\alpha\right)$ represents a potential tragedy of the commons. Although selecting an appropriate initial condition can stabilize the system at $(x_{2t}, 1, \alpha)$, thereby achieving coexistence between cooperators and defectors, the underlying risk remains at $r = 1$. This implies that if the collective threshold $M$ is not met, the group will inevitably fail and lose their endowments, even when a majority of individuals contribute the maximum cost $\alpha$. In addition, we find that the basin of attraction for this stable edge equilibrium point is larger than that of the stable corner equilibrium point (Fig B in S1 File).

Similarly, when the edge equilibrium points $(x_{1b}, 1, \beta)$ and $(x_{2b}, 1, \beta)$ exist, the equilibrium $(x_{2b}, 1, \beta)$ is stable if $\frac{\theta_1 + \theta_3}{\theta_1 + \theta_2} < x_{2b} < \frac{\mu_1}{\mu_1 + \mu_2}$. The validation of these findings is illustrated through numerical calculations in Fig~\ref{fig: model_1_figure1}f, \ref{fig: model_1_figure1}g and \ref{fig: model_1_figure1}h. Ultimately, the system exhibits bistable outcomes: depending on the initial conditions, trajectories converge either to the corner equilibrium $(0,1,\alpha)$ or to the edge equilibrium $(x_{2b}, 1, \beta)$. The former represents the worst-case scenario (full defection, maximal cost, and maximal risk). The latter, in contrast, depicts a state of stable coexistence. In this state, despite the persistent high risk ($r=1$), cooperators succeed in meeting the collective target $M$ and only bear the minimum cooperation cost $\beta$. This outcome provides a possible escape route from the tragedy of the commons, as achieving the threshold allows cooperators to retain their endowments despite the ever-present risk of failure. Furthermore, it is shown that the basin of attraction for the stable edge equilibrium point is larger than that of the corner equilibrium point (Fig B in S1 File). Compared to the scenario shown in Fig~\ref{fig: model_1_figure1}c, where cooperators bear the cost $\alpha$, the majority of cooperators here contribute only the minimum cost $\beta$.

Notably, both edge equilibrium points $(x_{2t}, 1, \alpha)$ and $(x_{2b}, 1, \beta)$ are stable when $x_{2t}<\frac{\theta_1 + \theta_3}{\theta_1 + \theta_2}<x_{2b}<\frac{\mu_1}{\mu_1 + \mu_2}$. This theoretical result is further supported by numerical calculations. Depending on the initial conditions, the system may converge to the corner equilibrium $(0,1,\alpha)$ or to one of the edge equilibria $(x_{2b}, 1, \beta)$ and $(x_{2t}, 1, \alpha)$. This tristability is shown in Fig~\ref{fig: model_1_figure1}i. The observed system behavior highlights the role of initial conditions in determining whether the tragedy of the commons can be avoided (see Fig~\ref{fig: model_1_figure1}j and \ref{fig: model_1_figure1}k).
\begin{figure*}
    \centering
    \includegraphics[width=\linewidth]{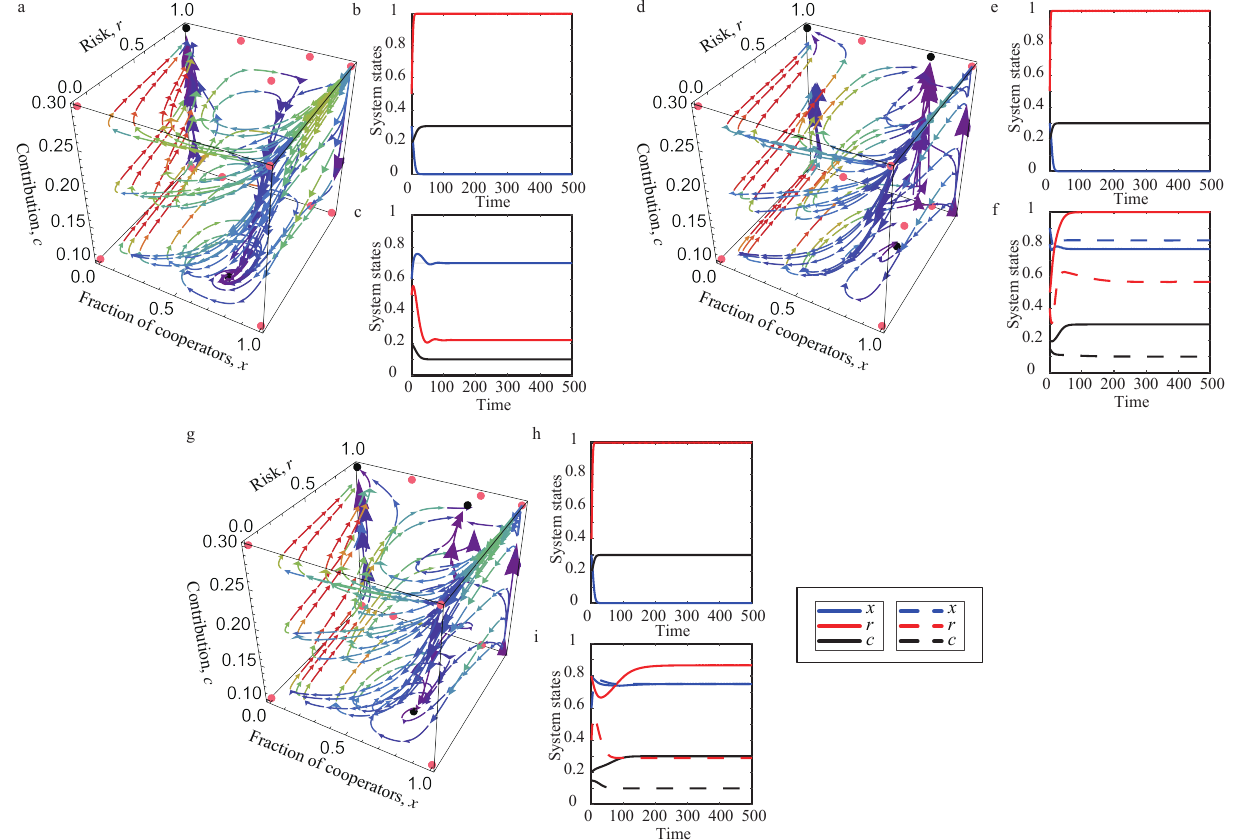}
    \caption{{\bf Bistable and tristable equilibria in coevolutionary dynamics.} Panels~a, d, and g show the phase flow in $x-r-c$ phase space when the system exhibits bistability and tristability, respectively. Panels~b, c, e, f, h, and i show the time evolution of the coevolving variables. Parameters are $\mu_1=0.7$, $\theta_3=0.2$ in panels~a, b, and c; $\mu_1=1.4$, $\theta_3=0.8$ in panels~d, e, and f; $\mu_1=0.9$, $\theta_3=0.8$ in panels~g, h, and i. The remaining parameters $N=8$, $M=5$, $\mu_2=0.3$, $\theta_1=0.5$, $\theta_2=0.5$, $\theta_4=0.6$, $b=2$, $\beta=0.1$, and $\alpha=0.3$ are fixed across all panels.}
    \label{fig: model_1_figure2}
\end{figure*}

We next examine the stability of surface equilibria in the phase space. The surface equilibrium $(R_x, 1, R_c)$ is unstable. Another surface equilibrium is $(x^\ast, T_r, \alpha)$, where the cost of cooperation is at its maximum ($c=\alpha$) and the risk satisfies $r=T_r\in(0,1)$. This state is stable if $T_1 < 0$ and $T_5 < 0$. Finally, the equilibrium $(x^\ast, B_r, \beta)$ lies on the $c=\beta$ surface, with the risk satisfying $r=B_r\in(0,1)$. This state is stable when $B_1 < 0$ and $B_5 < 0$.

Based on the above equilibrium analysis, we now present numerical examples to illustrate three representative multistable evolutionary outcomes. As depicted in Fig~\ref{fig: model_1_figure2}a, \ref{fig: model_1_figure2}b and \ref{fig: model_1_figure2}c,  the system exhibits bistability between the surface equilibrium $(x^\ast, B_r, \beta)$ and the corner equilibrium $(0,1,\alpha)$. This indicates that, depending on the initial state, the population may converge either to the tragedy of the commons or to a mixed state where cooperators coexist with defectors at risk level $B_r$. A tristable regime is illustrated in Fig~\ref{fig: model_1_figure2}d, \ref{fig: model_1_figure2}e and \ref{fig: model_1_figure2}f, where three stable equilibria coexist: the surface equilibrium $(x^\ast, B_r, \beta)$, the corner equilibrium $(0,1,\alpha)$, and the edge equilibrium $(x_{2t}, 1, \alpha)$.  In this case, cooperation can be sustained in two distinct ways: either at the minimum cost $\beta$ at the surface equilibrium or at the maximum cost $\alpha$ at the edge equilibrium with $r=1$. Fig~\ref{fig: model_1_figure2}g, \ref{fig: model_1_figure2}h and \ref{fig: model_1_figure2}i illustrate another example of tristability. In this regime, the corner equilibrium $\left(0,1,\alpha\right)$ and the two surface equilibria $\left(x^\ast,B_r,\beta\right)$ and $\left(x^\ast,T_r,\alpha\right)$ are all stable. Here, two alternative mixed states are possible: one with low cost $\beta$ and another with high cost $\alpha$, associated with risk levels $B_r$ and $T_r$, respectively. We note that the interior equilibrium point, if it exists, is unstable.

\begin{figure}
    \centering
    \includegraphics[width=\linewidth]{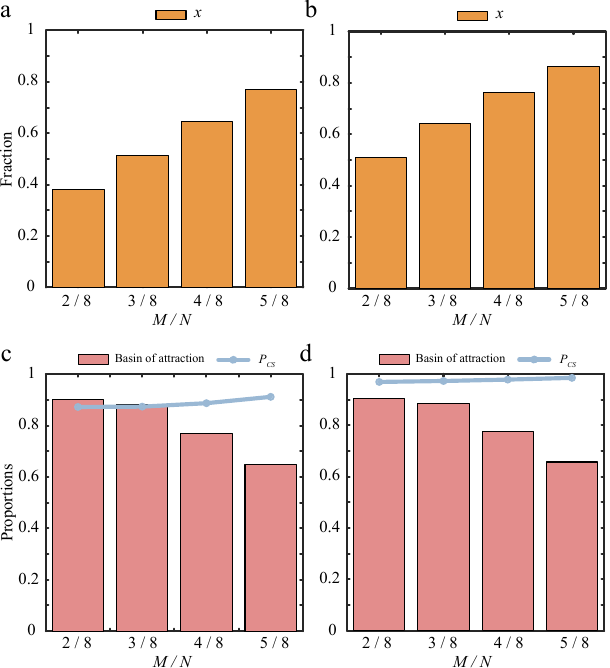}
    \caption{{\bf Impact of the threshold $M$ on the stability and success rate of collective action at two edge equilibrium points.} Panels~a and b show the fraction of cooperators (bars) corresponding to the stable edge equilibrium points $(x_{2t}, 1, \alpha)$ and $(x_{2b}, 1, \beta)$, respectively. Panels~c and d show the basin of attraction (bars) and the success rate of cooperation (lines) corresponding to the stable edge equilibrium points $(x_{2t}, 1, \alpha)$ and $(x_{2b}, 1, \beta)$, respectively. Parameters are $\theta_2=0.7$, $\theta_3=0.6$, $\theta_4=0.6$ in panels~a and c; $\theta_2=2$, $\theta_3=0.15$, $\theta_4=0.6$ in panels~b and d. The remaining parameters $N=8$, $\mu_1=2.5$, $\mu_2=0.3$, $\theta_1=0.5$, $b=2$, $\beta=0.1$, and $\alpha=0.3$ are fixed across all panels.}
    \label{fig:coverge}
\end{figure}

We next examine the effect of varying the collective threshold $M$ on cooperative outcomes at the two edge equilibria $(x_{2t}, 1, \alpha)$ and $(x_{2b}, 1, \beta)$. For the former equilibrium, our results are summarized in Fig~\ref{fig:coverge}a, which shows the dependence of the fraction of cooperators on the threshold $M$. We find that the fraction of cooperators increases monotonically with $M$. Similar results are obtained when the system stabilizes at the other edge equilibrium $(x_{2b}, 1, \beta)$, as shown in Fig~\ref{fig:coverge}b. For the first equilibrium, Fig~\ref{fig:coverge}c shows how, as $M$ increases, the size of the basin of attraction and the success rate of collective action vary. The corresponding plot for the equilibrium $(x_{2b}, 1, \beta)$ is shown in Fig~\ref{fig:coverge}d. Both panels show that the basin of attraction gradually shrinks, whereas the success rate of collective action increases slightly as the threshold $M$ is raised. These results indicate that raising the collective threshold can enhance the level of cooperation under stable conditions and promote the success of collective action.

\section*{Discussion}
Human behavior and ecosystems are intertwined through complex and profound interactions in modern societies~\cite{perc2017statistical,liu2007complexity,farahbakhsh2022modelling}. Recent theoretical research has focused on feedback mechanisms in collective-risk dilemmas, highlighting bidirectional coupling between individual decisions and collective risk, as well as between decisions and the cost of cooperation. However, a more complex tripartite feedback among cooperation levels, collective risk, and the cost of cooperation remains underexplored in the existing literature~\cite{liu2023coevolutionary,hua2024coevolutionary2}. In this work, we develop a feedback-evolving game model to capture these coupled dynamics. Specifically, we assume that an increase in the fraction of cooperators reduces both collective risk and the cost of cooperation, whereas an increase in collective risk raises the cost of cooperation.

Our findings indicate that the tragedy of the commons equilibrium $(0,1,\alpha)$, characterized by full defection, maximal risk, and maximal cooperation cost, is always stable, implying a persistent risk of collective failure. However, the system also exhibits multistability. The system can also stabilize at edge equilibria such as $(x_{1b},1,\beta)$ or $(x_{1t},1,\alpha)$, where cooperators coexist with defectors under high risk while paying either minimal or maximal cooperation costs. Furthermore, surface equilibria like $(x^\ast,T_r,\alpha)$ and $(x^\ast,B_r,\beta)$ represent states where cooperation is maintained at constant high or low cost, with an intermediate stationary risk level. Critically, the final equilibrium depends on the initial conditions. This sensitivity to initial conditions implies that early-stage interventions affecting cooperation levels or the cost of cooperation may play a decisive role in driving the system toward more desired cooperative outcomes.
 
Our base model (Model~1) incorporates feedback from cooperation to risk, but does not include feedback from cooperation cost to risk. To explore this additional coupling, we introduce an extended model (Model 2) in the S1 File. In Model 2, an increase in cooperation or cooperation cost reduces risk, while an increase in defection or risk elevates this cost (Fig C in S1 File). By analyzing the modified replicator system, we have identified several novel dynamics. These include the monostable state $(0,0,\alpha)$, which depicts the disappearance of cooperation as the highest cooperation cost is required in a risk-free scenario (Fig D in S1 File). We also observe oscillatory dynamics of risk and cooperation cost while the population remains in a full defection state (Fig F in S1 File). Furthermore, the system can exhibit tristable equilibrium points (Fig F in S1 File). From a policy perspective, these results suggest that temporarily lowering the cost of cooperation, for example through subsidies or incentive mechanisms introduced by third-party institutions, may help the population move away from unfavorable states and reach cooperative equilibria, even if such interventions are not permanent.

Finally, we note that our study has certain limitations. As a first approach, we assumed a linear feedback relationship between cooperation level, cooperation cost, and collective risk, while the feedback relationships between population behaviors and the social environment could be more subtle~\cite{weitz2016oscillating,milinski2008collective,cronk2021design,tilman2020evolutionary}. Furthermore, our analysis relies on the infinite well-mixed population assumption, which neglects stochastic factors. Future research could extend our framework by incorporating nonlinear functional forms and exploring coevolutionary dynamics in finite or structured populations.

\section*{Supporting information} S1 File. This file contains eight figures and one table, presenting theoretical analyses and robustness investigations of two distinct feedback game models. Figs A and B illustrate the robustness results of Model 1. Figures C through H display the theoretical and numerical results of Model 2. Table A summarizes the stability conditions of the equilibrium points in Model 2.

\section*{Data availability statement}
Variants of multiple feedback game models presented in supplementary information. Codes used to generate the results in the manuscript and in the supplementary information are available at GitHub: \url{https://github.com/lichengwanw/collective-risk-games}.

\section*{Author Contributions}
\noindent\textbf{Conceptualization:} Lichen Wang, Shijia Hua\\
\noindent\textbf{Data curation:} Lichen Wang, Shijia Hua\\
\noindent\textbf{Formal analysis:} Lichen Wang, Shijia Hua, Yuyuan Liu \\
\noindent\textbf{Funding acquisition:} Shijia Hua, Linjie Liu \\
\noindent\textbf{Investigation:} Lichen Wang, Shijia Hua \\
\noindent\textbf{Methodology:} Lichen Wang, Shijia Hua, Yuyuan Liu\\
\noindent\textbf{Project administration:} Lichen Wang, Linjie Liu\\
\noindent\textbf{Resources:} Linjie Liu\\
\noindent\textbf{Software:} Lichen Wang, Yuyuan Liu\\
\noindent\textbf{Validation:} Linjie Liu\\
\noindent\textbf{Writing - original draft:} Lichen Wang, Shijia Hua\\
\noindent\textbf{Writing - review \& editing:} Shijia Hua, Linjie Liu, Attila Szolnoki, Liang Zhang


\section*{Competing interests}
The authors have declared that no competing interests exist
%
%

\bibliography{plos_bibtex_sample}

\end{document}